\documentclass[prd,twocolumn,superscriptaddress,altaffilletter,showpacs,nofootinbib]{revtex4}

\usepackage{amssymb}
\usepackage[dvips]{graphicx}
\usepackage{amsmath}

\newcommand{\be}{\begin{equation}}
\newcommand{\ee}{\end{equation}}
\newcommand{\bea}{\begin{eqnarray}}
\newcommand{\eea}{\end{eqnarray}}

\newcommand{\vphi}{\varphi}

\begin{document}

\title{Gaussian Warp Factor: Towards a Probabilistic Interpretation of Braneworlds}

\author{Israel Quiros}\email{iquiros6403@gmail.com}\affiliation{Departamento de Matem\'aticas, Centro Universitario de Ciencias Ex\'actas e Ingenier\'{\i}as, Corregidora 500 S.R., Universidad de Guadalajara, 44420 Guadalajara, Jalisco, M\'exico.}

\author{Tonatiuh Matos}\email{tmatos@fis.cinvestav.mx}\affiliation{Departamento de F\'{\i}sica, Centro de Investigaci\'on y de Estudios Avanzados del I. P. N. Apdo. Post. 14-740, 07000, M\'exico, D.F., M\'exico}

\date{\today}

\begin{abstract}
We investigate Gaussian warped five-dimensional thick braneworlds. Identification of the graviton's wave function (squared) in the extra-dimension with a probability distribution function leads to a straightforward probabilistic interpretation of braneworlds. The extra-coordinate $y$ is regarded as a Gaussian-distributed random variable. Hence, all of the field variables and operators which depend on $y$ are, also, randomly distributed. Four-dimensional measurable (macroscopic) quantities are identified with the corresponding averaged values over the Gaussian distribution. The present scenario represents a new phenomenological approach to smooth thick branes which can not be obtained through 'smearing out' Randall-Sundrum-like (thin) braneworlds.
\end{abstract}

\pacs{04.50.-h, 04.50.Cd, 11.25.Mj, 11.25.-w, 04.20.Jb, 02.50.-r}

\maketitle

\section{Introduction}\label{intro}

Braneworld models \cite{arkani-hamed}--\cite{tanaka} represent an interesting alternative to Kaluza-Klein (KK) compactification even if the extra-space can be of infinite extension \cite{rs}. In spite of the success of braneworlds models with a 0-width in the extra-space, at least at string scale the thickness of the brane cannot be neglected. The thin-brane approximation is valid as long as the explored energy scales are much smaller than the inverse thickness of the brane. In \cite{arkani-hamed}, for instance, the width of the brane was assumed to be of the order of the distance at which the electroweak interactions have been probed $\sim m^{-1}_{EW}$ ($m_{EW}\sim 10^3$ GeV is the electro-weak energy scale). These more realistic alternatives to 0-width brane configurations are known as thick braneworlds \cite{gremm}--\cite{arias}. A problem with these models is what to identify as a four-dimensional observable quantity. In cosmological applications, for instance, due to the finite thickness of the brane, there is some arbitrariness in the definition of what the effective 4D quantities should be. The simplest prescription one can envisage is to define the 4D effective quantity associated to a 5D quantity as its spatial average over the brane thickness \cite{mounaix} (see also \cite{quiros-matos}). Needless to say that the above prescription is legitimate only if we deal with branes with well-defined thickness. However, in the case when the brane is smoothly spread over the extra-space, there is not a unique way to choose the brane thickness, and this leads to ambiguities in the computation of 4D measurable quantities.

In the present paper we shall propose an alternative way of computing effective quantities within 5D smooth thick brane contexts without ambiguities, even if the brane thickness can not be chosen in a unique way. Our approach will rely on the following assumptions: i) the background geometry is a warped one with the warp factor being a Gaussian in the extra-coordinate, and ii) the extra-coordinate is to be regarded as a continuous random parameter with a Gaussian probability density. The resulting picture amounts to a probabilistic representation of braneworlds. In this approach the large hierarchy between the TeV and Plack scales may be explained if assume an additional chameleonic interaction between the matter degrees of freedom living in the brane and the 5D scalar field.

\section{The Model}\label{model}

In this paper we shall explore thick braneworlds supported by a scalar field. The thick branes are described by the following 5D action: 

\bea S_5=\int d^5x\sqrt{|g_5|}\left[\frac{R_5}{2\kappa^2_5}-\frac{1}{2}(\nabla_5\vphi)^2-V(\vphi)\right],\label{action}\eea where $\kappa^2_5 \simeq 1/M_5^3$ ($M_5$ is the 5D Planck mass), and $V(\vphi)$ is the scalar field's ($\vphi$) self-interaction potential. Besides, $(\nabla_5\vphi)^2=g^{MN}\vphi_{,M}\vphi_{,N}$ , $\Box_5\vphi=g^{MN}\vphi_{;MN}$. Here capital Latin indexes $A,B,...,M,N...=0,1,2,3,5$, while small Greek indexes $\alpha,\beta,...,\mu,\nu,...=0,1,2,3$. The derived 5D field equations are the Einstein's field equations (EFE), and the Klein-Gordon equation (KGE) for the scalar field:

\be G_{AB}=\kappa_5^2\,T^{(\vphi)}_{AB},\;\Box_5\vphi=dV/d\vphi,\label{efe}\ee respectively, where $G_{AB}\equiv R_{AB}-g_{AB}R_5/2$ is the 5D Einstein's tensor, and

\be T^{(\vphi)}_{AB}=\vphi_{,A}\vphi_{,B}-\frac{1}{2}g_{AB}(\nabla_5\vphi)^2-g_{AB}V,\label{phi-stress-energy-t}\ee is the scalar field's stress-energy tensor. 

To start with, let us to assume a regular 5D background metric which respects 4D Poincar\`e invariance:

\be ds_5^2=a^2(y)\eta_{\mu\nu}dx^\mu dx^\nu+dy^2,\label{metric}\ee where $\eta_{\alpha\beta}$ is the 4D Minkowski metric, $a^2(y)$ is the warp factor, and, for the moment being, only dependence on the extra-coordinate $y$ is being considered. For definiteness we will assume that all of the coordinates ${\bf x}\equiv x^\mu$, $y$, are measured in the same length units, while the warp factor $a^2$ is dimensionless. In view of (\ref{metric}), the field equations (\ref{efe}) can be explicitly written in the following way:

\bea &&3H'+6H^2=-\frac{\kappa_5^2}{2}\left(\vphi'^2+2V\right),\nonumber\\
&&6H^2=\frac{\kappa_5^2}{2}\left(\vphi'^2-2V\right),\;\vphi''+4H\vphi'=dV/d\vphi,\label{feqs}\eea where $H\equiv a'/a$, and the tilde denotes derivative with respect to the extra-coordinate $y$. Only two of the above equations are independent from each other.

Due to the additional degrees of freedom associated with the scalar field (and its self-interaction potential) one may choose a suitable functional form of the warp factor from the start and, then, one solves the equations that involve the scalar field and the self-interaction potential. In the present case, in order to derive exact solutions, we make the following (Gaussian) ansatz for the warp factor:

\be a^2(y)=\exp{(-\mu^2 y^2)},\label{warp-f}\ee where the parameter $\mu$ is the inverse of the brane thickness $\Delta=1/\mu$. As long as $a^2(y)=a^2(-y)$, here we shall be focusing in 5D solutions which respect mirror symmetry: $\vphi(y)=\pm\vphi(-y)$, $V(\vphi)=V(-\vphi)$. 

We can combine equations (\ref{feqs}) into derived equations that can be easier to handle: $3H'=-\kappa_5^2\vphi'^2$ $\Rightarrow$ $\vphi(y)=\pm\sqrt{3\mu^2/\kappa_5^2}\;y+C$, where the ''$\pm$'' signs account for two possible branches of the solution, and $C$ is an integration constant which has to be set to zero if one requires mirror symmetry to be a symmetry of the solution. Besides, the second equation in (\ref{feqs}) can be rewritten as, $\kappa_5^2 V=\kappa_5^2\vphi'^2/2-6H^2$, i. e., $V=3\mu^2(1-4\mu^2 y^2)/2\kappa_5^2$. Hence, we obtain the following solution for the scalar field $\vphi$ and the potential $V(\vphi)$:

\be \vphi(y)=\pm\sqrt{3\mu^2/\kappa_5^2}\;y,\;V(\vphi)=V_0-2\mu^2\,\vphi^2,\label{5d-sol}\ee where $V_0=3\mu^2/2\kappa_5^2$. This solution is characterized by diverging values of the curvature invariants as $y\rightarrow\infty$. Nevertheless, as it will be discussed in section \ref{gaussian-distribution}, this does not place any problems to the picture we are about to expose. 

The potential $V(\vphi)$ in (\ref{5d-sol}) makes the given scalar field's solution unstable. This has no consequences as long as $\vphi=\vphi(y)$ and minimal coupling of matter is considered, since the 4D gravity and the evolution of the scalar field are decoupled in this case. However, if consider additional non-gravitational interaction of matter with $\vphi$, the instability induced by the above potential can have catastrophic consequences. Nonetheless, if assume $\vphi=\vphi({\bf x},y)$ from the start, this problem is overcame as a result of the chameleon effect (see section \ref{matter}).

A few additional comments on our solution. The RS solution \cite{rs2} can not be recovered from (\ref{warp-f},\ref{5d-sol}) in the limit $\Delta\rightarrow 0$ ($\mu\rightarrow\infty$). Hence, our model does not represent a thick brane generalization of the RS model as it is the case for a large body of thick brane scenarios found in the bibliography \cite{csaba,arias,polacos}. It is just a different phenomenological scenario. Notice that $\vphi(y)$ in (\ref{5d-sol}) has not a kink-like profile as it is customary in the bibliography on smooth thick branes made out of a scalar field (see, for instance, Ref.\cite{polacos}). The fact that the solution (\ref{warp-f},\ref{5d-sol}) supports a smooth thick brane configuration will be evident below when we study the gravitational content of our model. It will be seen that a massless bound gravitational state does actually exists which is localized around the origin of the extra-coordinate. The smooth thick brane is shaped in the extra-space by the wave-function (squared) of this gravitational state which can be identified with the 4D graviton.

\section{Graviton's Wave Function}\label{wave-function}

Here we shall consider linear perturbations of the metric (\ref{metric}): $ds_{5,pert}^2=a^2(y)\left[\eta_{\mu\nu}+\epsilon\,h_{\mu\nu}({\bf x},y)\right] dx^\mu dx^\nu+dy^2$, where $\epsilon\,h_{\mu\nu}({\bf x},y)$ are small perturbations around the Minkowski metric. We shall consider the transverse and traceless (TT) gauge, $h=\eta^{\mu\nu}h_{\mu\nu}=\eta^{\mu\nu}h_{\alpha\mu,\nu}=0$, since, in this gauge, the linear perturbations of the metric decouple from perturbations of the scalar field \cite{dwolfe}. The linearly perturbed EFE-s read: $a^{-2}\Box_{(\eta)}h_{\mu\nu}+h''_{\mu\nu}+4H h'_{\mu\nu}-2(3H'+8H^2)h_{\mu\nu}=\kappa_5^2(\vphi'^2+2V)h_{\mu\nu}$, or, if make the ansatz $h_{\mu\nu}({\bf x},y)=a^{-2}(y)\,\Psi(y)\,\pi_{\mu\nu}({\bf x})$, where $\Psi(y)$ is the wave function of the gravitational modes in the extra-dimension, then the latter equation can be expressed in the form of a Schroedinger equation,

\be \left[-\frac{\partial^2}{\partial y^2}+U_{QM}(y)\right]\Psi(y)=m^2\,a^{-2}\,\Psi(y),\label{schroedinger-eq}\ee where $U_{QM}(y)=-2\mu^2(1-2\mu^2 y^2)$ is the Schroedinger-like quantum mechanical potential, and $m^2$ accounts for the 4D mass of the excited KK gravitational modes: $\Box_{(\eta)}\pi_{\mu\nu}({\bf x})=m^2\pi_{\mu\nu}({\bf x})$. Equation (\ref{schroedinger-eq}) can be explicitly written as

\be \Psi''+2\mu^2(1-2\mu^2y^2)\Psi=-m^2\sqrt N.\label{wave-eq}\ee 

For the massless mode (properly the 4D graviton): $\Psi_0''+2\mu^2(1-2\mu^2 y^2)\Psi_0=0$, which solution is $\Psi_0(y)=\sqrt N\exp{(-\mu^2y^2)}$, where, if normalize the wave function $\Psi_0$ to unity, $\int_{-\infty}^\infty\Psi^2_0(y)\,dy=1$ $\Rightarrow\;N=\sqrt{2/\pi}\,\mu$. The solution of Eq.(\ref{wave-eq}) for the continuum of KK states is given by: $\Psi_m(y)=\sqrt{\pi/2}\,(m/2\mu)^2\Psi_0(y)\,E(y)$, where $$E(y)\equiv-\text{erf}(\mu y)\,\text{erfi}(\sqrt 2\mu y)+\sqrt\frac{2}{\pi}\int_0^{\sqrt 2\mu y} e^{-\frac{\xi^2}{2}}\,\text{erfi}(\xi)\,d\xi,$$ and, $\text{erf}(x)=2\int_0^x e^{-\xi^2}\,d\xi/\sqrt\pi$, $\text{erfi}(x)=2\int_0^x e^{\xi^2}\,d\xi/\sqrt\pi$, are the error function, and the imaginary error function respectively. 

\begin{figure}[t]
\includegraphics[width=3.5cm,height=3cm]{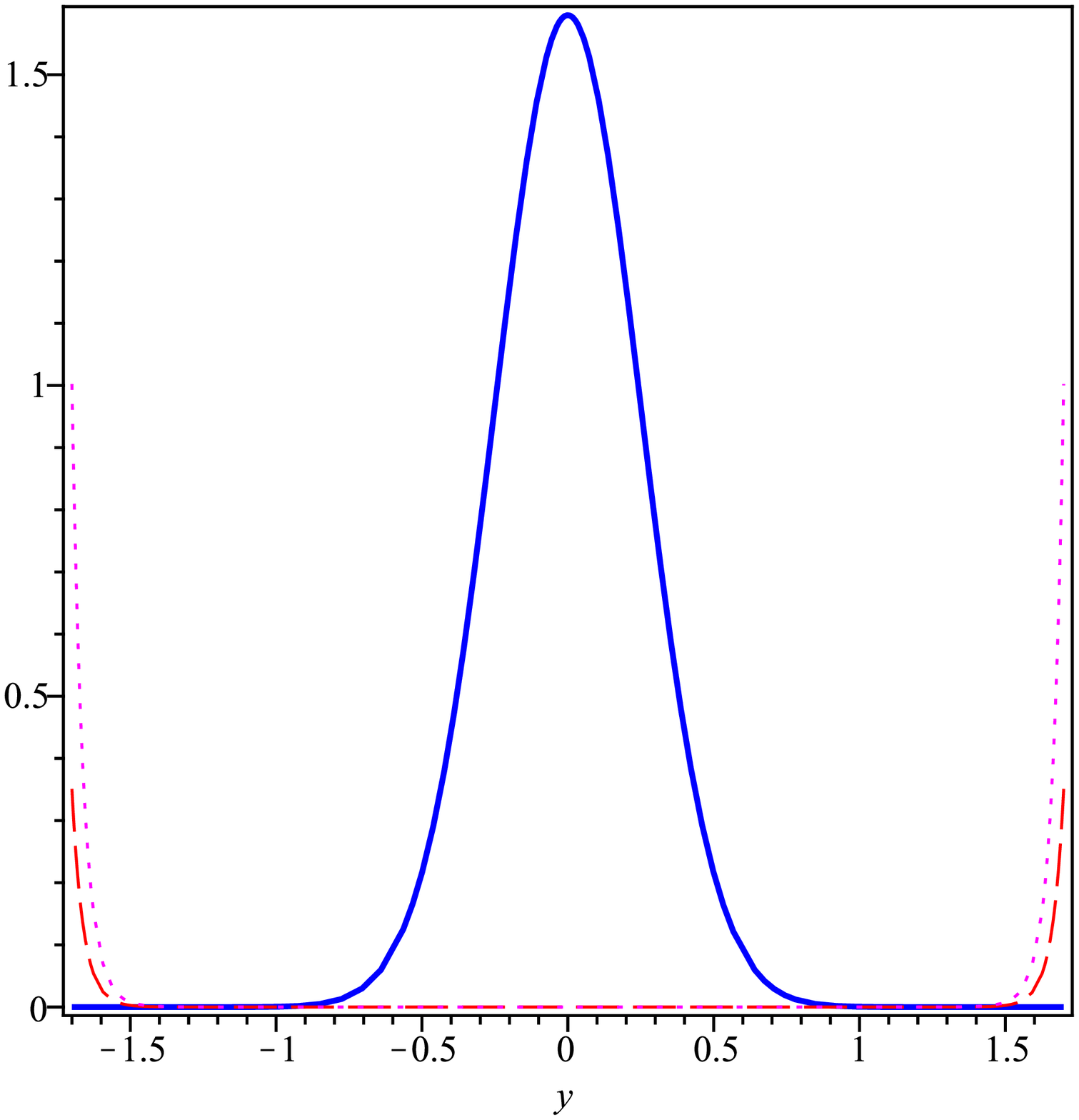}
\includegraphics[width=3.5cm,height=3cm]{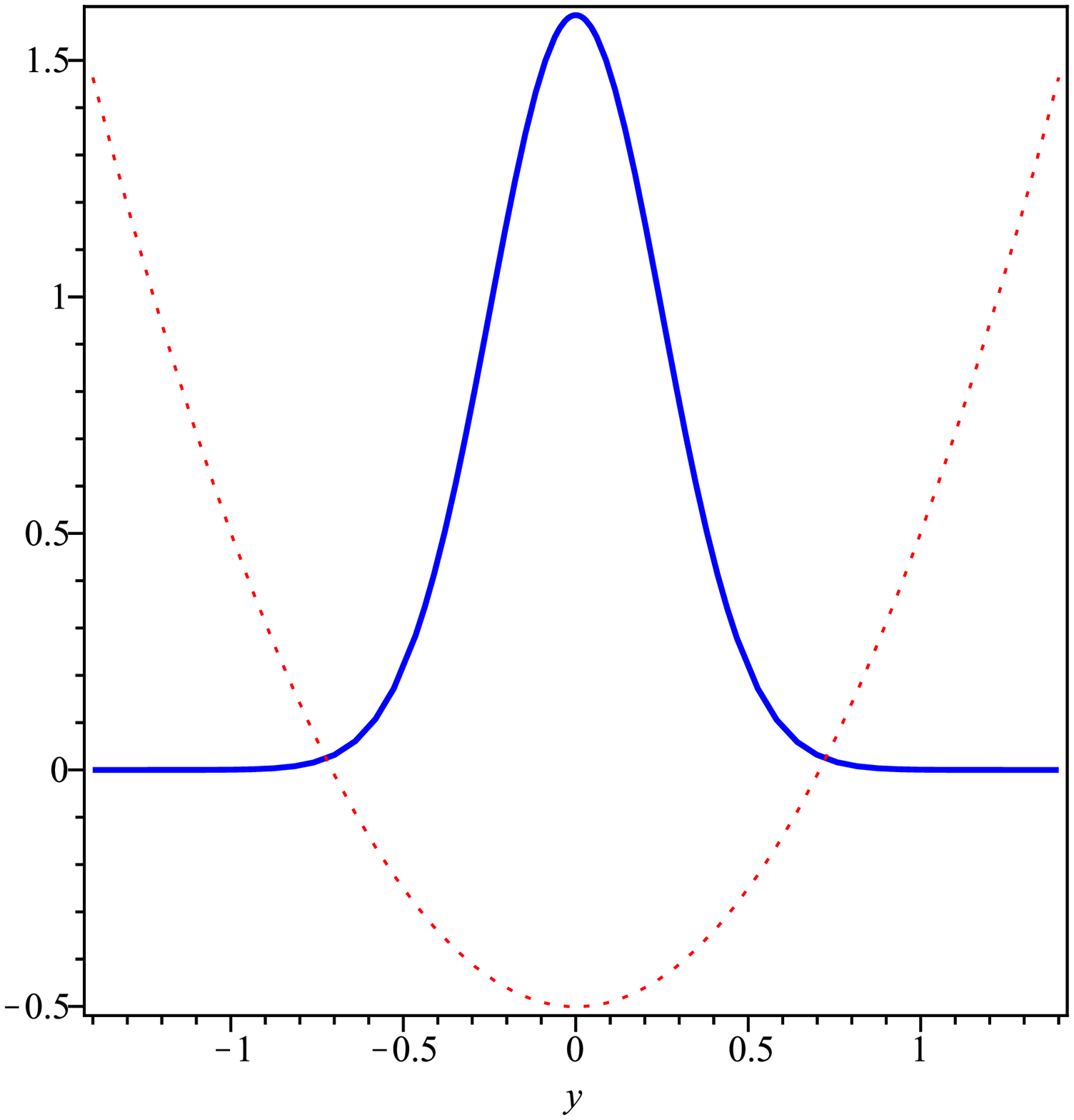}
\caption{Probability density profile for the gravitational states. The 4D graviton's probability density $\Psi^2_0(y)$ (solid line), and the probability density for the continuum of massive states $\Psi^2_m(y)$ (mass parameter $m=0.01$- dashed curve, and $m=0.013$ - dotted line) are shown in the left-hand figure. The brane thickness has been arbitrarily set $\Delta=0.5$. It is seen that the lightest mode of mass $m=0.01$ is localized farther away from the brane. In the right-hand figure the 4D graviton's probability $\Psi^2_0(y)$ is shown together with the Schroedinger-like potential $U_{QM}(y)$ times $10^{-1}$ (doted curve).}
\label{fig1}
\end{figure}

In the figure \ref{fig1} we show the profiles of the probability densities of the graviton $\Psi_0^2(y)$ and of the massive states $\Psi_m^2(y)$. The massless graviton is localized on the brane since the analog quantum mechanical potential $U_{QM}$ is well-shaped with a negative minimum inside the brane \cite{csaba}. It is seen that the continuum of massive states is localized away from the brane which is shaped by the massless graviton's wave-function. Besides, the lightest of the two massive gravitational modes shown (mass parameter $m=0.01$) is localized farther away from the brane than the heaviest one ($m=0.013$), i. e., heavier modes are closer to the brane.

\subsection{Corrections to the Newtonian Potential}

The $y$-depending gravitational potential between two point-like sources of masses $m_1$ and $m_2$ respectively, which are located at $y$, arises from contributions of the massless graviton $\Psi_0$ (standard piece), and from the continuum of massive states $\Psi_m$ (corrections):\footnote{This equation is a slight modification of the one in Ref.\cite{csaba}, to include the possibility for the point masses to be located at different positions within the thick brane.} $$v(r,y)\sim\frac{\kappa_5^2 m_1 m_2}{r}\Psi_0^2+\kappa_5^2\,\int_\epsilon^\infty\frac{m_1 m_2\,e^{-mr}}{r}\,\Psi_m^2\,dm,$$ where in the second term in the right-hand-side (RHS) one has to go to the limit $\epsilon\rightarrow 0$. After integrating within the correction's term we get $$v(r,y)\sim\frac{\kappa_5^2 m_1m_2}{r}\,\Psi_0^2(y)+\frac{3\pi\kappa_5^2 m_1m_2}{4\mu^4\,r^6}\,\Psi_0^2(y)E^2(y).$$ 

Next we integrate over an effective width of the brane $\tau$ to get a $y$-independent gravitational potential: 

\be V_N(r)\sim\frac{G_N m_1 m_2}{r}\left(1+\frac{3\pi\gamma}{4\mu^4}\frac{1}{r^5}\right),\label{4d-grav}\ee where $G_N=\kappa_5^2\int_{-\tau}^\tau \Psi_0^2(y)\,dy$, and $$\gamma=\int_{-\tau}^\tau\Psi_0^2(y)\,E^2(y)\,dy/\int_{-\tau}^\tau\Psi_0^2(y)\,dy,$$ is a small constant since the massive gravitational states are localized away from the brane as seen in the figure \ref{fig1}.  

We have seen that there is a normalizable bound zero-mass gravitational state whose wave-function shapes the form of the brane in the extra-space. The continuum of massive KK modes produces only very small (negligible) corrections to the Newtonian gravitational potential which fall-off very quickly as $\propto r^{-6}$. This is expected since the analog quantum mechanical potential $U_{QM}(y\rightarrow\pm\infty)>0$ (in fact it uncontrollably grows up at large $|y|$-s). In this case the excited KK-states are separated by a gap from the ground state \cite{csaba}. In our set-up this may be understood in the following way. The massive modes are localized away from the brane: the lighter ones are farther away while the heavier KK states are closer to the brane. Hence, the lighter states may not be excited neither by low-energy processes living in the thick brane, nor by self-gravitating interactions with the bound graviton, since the probability to find the latter far from the brane goes like $\propto\exp{(-2\mu^2y^2)}$ and vanishes very quickly. Only heavier KK modes which are close enough to the thick brane might by excited by interactions with the graviton, but this would need of an energy of the order of the KK mode's mass, and so are effectively decoupled from the thick brane also. 

A crude estimate of the effective mass gap $m_{eff}$ can be given if make the following reasonable assumptions: i) there is a position $y_*$ in the extra-space where the contributions from the massless and excited KK modes are of the same order $\Psi_0^2(y_*)\simeq\Psi_m^2(y_*)$, and ii) $E^2(y_*)$ is of order unity. Under these assumptions $m_{eff}\sim 2(2/\pi)^{1/4}\mu\sim(2/\pi)^{1/4}2/\Delta$ (recall that $\Delta$ is the brane thickness). Hence, for sharper probability densities the effective mass gap is larger. If assume, for instance, that the width of the brane is of the order of the distance at which electroweak interactions have been probed $\sim m_{EW}^{-1}$, then the effective mass gap is of the order of the electroweak energy scale $\sim m_{EW}\sim 10^3$ GeV.

\section{Gaussian Probability Distribution}\label{gaussian-distribution}

After the Gaussian ansatz (\ref{warp-f}) for the warp factor, the next step is to regard the extra-coordinate $y$ as a random variable which follows a normal (Gaussian) distribution centered at the origin $y=0$ with probability density function $$f(y)=\frac{1}{\sqrt{2\pi}\,\sigma}\,e^{-\frac{y^2}{2\sigma^2}},\;\int_{-\infty}^{\infty}f(y)\,dy=1,$$ where $\sigma^2$ is the standard deviation. Since the massless gravitational state's wave-function $\Psi_0(y)$ shapes the profile of the brane in the extra-dimension, it seems quite reasonable to identify the graviton's probability density $\Psi^2_0(y)$ with the probability density function for the random variable $y$: 

\be f(y)\;\rightarrow\;\Psi^2_0(y)=\sqrt\frac{2}{\pi}\frac{1}{\Delta}\,e^{-\frac{2y^2}{\Delta^2}},\label{gauss-d}\ee where, for definiteness, we identify $2\sigma$ with the brane's width $\Delta=1/\mu$. This way one endows the braneworld picture with a probabilistic interpretation where all of the field variables which depend on $y$ behave randomly and are Gaussian-distributed quantities. Besides, the probability distribution function $\Psi^2_0(y)$ induces a ''natural'' prescription for what to identify as a four-dimensional observable within thick brane contexts. Actually, in the present case a 4D measurable (effective) quantity associated to a 5D quantity $Q({\bf x},y)$ coincides with its averaged value over the Gaussian distribution: $\left\langle Q({\bf x},y)\right\rangle=\int_{-\infty}^{\infty} Q({\bf x},y)\,\Psi^2_0(y)\,dy$.\footnote{Integrating over the infinite extent of the extra-space is legitimate only for processes which occur at energies below the scale at which the KK modes couple effectively to the 4D graviton. Otherwise one should integrate over an effective brane thickness.}  

For illustration let us to compute the averaged 5D Ricci curvature: $\left\langle R_5\right\rangle=3\mu^2$. Hence, even if $R_5$ diverges at $y$-infinity as $\propto-y^2$, its averaged value $\left\langle R_5\right\rangle$ -- the one measured by a 4D observer -- is a finite quantity. This means that the 5D gravity, as understood by a 4D observer, is asymptotically de Sitter.

\subsection{Macroscopic Quantities and Probabilities}

Under the above prescription for a measurable 4D quantity, the thick brane picture acquires a simple probabilistic interpretation. Given that the extra-coordinate $y$ is regarded as a random variable, then, the $y$-dependent gravitational and matter fields are submitted to 'microscopic' random field variables $\mu_i({\bf x},y)$, such as the metric $g_{\mu\nu}({\bf x},y)$ (and derived curvature objects), the stress-energy tensor of matter $T_{\mu\nu}^{(m)}({\bf x},y)$, etc. Their corresponding 'macroscopic' (measurable) quantities $M_i({\bf x})$, are computed through Gaussian averaging: 

\be M_i({\bf x})\equiv\left\langle\mu_i({\bf x},y)\right\rangle=\int_{-\infty}^{\infty} \mu_i({\bf x},y)\,\Psi^2_0(y)\,dy.\label{average}\ee 

One can average microscopic operators $\hat l_i$, equations, etc., and, in particular, the average of the microscopic field equations $e_i=0$, leads to the corresponding macroscopic equations, i. e., those which represent 4D physical laws testable by a 4D observer like us: $E_i=\left\langle e_i\right\rangle=0$.

One can also introduce probabilities for measured quantities as it follows. The 'probability' ${\cal P}$ of measuring $\mu_i$ within the $y$-interval $[y,y+\delta y]$ around the position $y=y_0$, is proportional to $\left({\cal P}_{\mu_i}\right)_{y_0}\propto\,\mu_i({\bf x},y_0)\,\Psi^2_0(y_0)\,\delta y=\sqrt{2/\pi}\exp{(-2y_0^2/\Delta^2)}\,\mu_i({\bf x},y_0)\,\delta y/\Delta$. In particular, the ratio of the probabilities to measure $\mu_i$ around $y=\pm y_0$ and around the origin $y=0$ (same interval $\delta y$) is $\left({\cal P}_{\mu_i}\right)_{y_0}/\left({\cal P}_{\mu_i}\right)_0=\exp{(-2y_0^2/\Delta^2)}\,\mu_i({\bf x},y_0)/\mu_i({\bf x},0)$.

\section{Effective 4D Gravity}\label{eff-4d-grav}

In standard Kaluza-Klein \cite{kk,wands} as in the alternative Randall-Sundrum (RS) compactification schemes \cite{rs}, after ensuring that a bounded massless graviton exists and can account for recovering of Newtonian gravity, one needs also to get a 4D effective description of the laws of gravity. This is usually done by straightforwardly integrating in respect to the extra-coordinate(s) under the action integral over the extension of the extra-dimensional manifold. In this paper we propose an alternative scheme. Here a bounded graviton exists by the same reason it does in RS compactification \cite{rs}: because of the warped property of the metric (\ref{metric}). However, unlike in standard KK and RS procedures, in our approach one does not integrate over $y$ within the action (\ref{action}) to get to an effective 4D picture. Instead, one regards the 5D microscopic field equations (\ref{efe}) as fundamental, and then, the macroscopic equations -- those which describe the 4D laws of gravity -- are obtained by computing the Gaussian average of (\ref{efe}). In this and the following sections we shall discuss our approach in more detail.     

Let us consider the metric given by the line element

\be ds_5^2 =a^2(y)\tilde{g}_{\mu\nu}({\bf x})dx^\mu dx^\nu+dy^2, \label{metric'}\ee  with the Gaussian warp factor defined as in (\ref{warp-f}), where in the line element (\ref{metric}) we have replaced the Minkowski metric by an arbitrary metric, $\eta_{\mu\nu}\rightarrow\tilde{g}_{\mu\nu}({\bf x})$. In this case the 5D 'microscopic' Einstein's field equations plus Klein-Gordon equation (\ref{efe}), can be written as:

\bea &&\tilde G_{\mu\nu}=\kappa_5^2\left[\vphi_{,\mu}\vphi_{,\nu}-\frac{1}{2}\tilde g_{\mu\nu}(\tilde\nabla\vphi)^2\right]-\nonumber\\
&&\;\;\;\;\;\;\;\;\;\;\;\;a^2\left[3H'+6H^2+\frac{\kappa_5^2}{2}(\vphi'^2+2V)\right]\tilde g_{\mu\nu},\nonumber\\
&&\frac{1}{2a^2}\left[\tilde R-\kappa_5^2(\tilde\nabla\vphi)^2\right]=6H^2-\frac{\kappa_5^2}{2}(\vphi'^2-2V),\nonumber\\
&&0=\kappa_5^2\vphi'\vphi_{,\mu}\,,\;a^{-2}\tilde\Box\vphi+\vphi''+4H\vphi'=dV/d\vphi,\label{4d-feqs}\eea where, as before, $H=a'/a$, and the quantities $\tilde Q_i$ are defined in terms of the 4D metric $\tilde g_{\mu\nu}$: $\tilde G_{\mu\nu}=\tilde R_{\mu\nu}-\tilde g_{\mu\nu}\tilde R/2$, $\tilde R=\tilde g^{\mu\nu}\tilde R_{\mu\nu}$, $(\tilde\nabla\vphi)^2=\tilde g^{\mu\nu}\vphi_{,\mu}\vphi_{,\nu}$, $\tilde\Box\vphi=\tilde g^{\mu\nu}\vphi_{;\mu\nu}$.\footnote{Needless to say that the semicolon in this last equation refers to covariant derivative in respect to the 4D metric $\tilde g_{\mu\nu}$.} The second equation in (\ref{4d-feqs}) comes from $G_{5\mu}=R_{5\mu}=0$. This equality forces $\vphi({\bf x})=\vphi_0$, i. e., the scalar field can not depend on the 4D spacetime point. Without loss of generality we can choose $\vphi=\vphi(y)$. After this, and recalling that $a^2(y)$ is given by the Gaussian function (\ref{warp-f}), the microscopic field equations are: 

\bea &&-\frac{\tilde G_{\mu\nu}}{a^2}=\left[3H'+6H^2+\frac{\kappa_5^2}{2}(\vphi'^2+2V)\right]\tilde g_{\mu\nu},\nonumber\\
&&\frac{\tilde R}{2a^2}=6H^2-\frac{\kappa_5^2}{2}(\vphi'^2-2V),\;\vphi''+4H\vphi'=\frac{dV}{d\vphi}.\nonumber\eea 

One can see that the solution (\ref{warp-f},\ref{5d-sol}) amounts to the macroscopic 4D vacuum EFE-s $\tilde G_{\mu\nu}=0$ ($\tilde R=0$). Where the average of the Einstein's tensor $\left\langle\tilde G_{\mu\nu}\right\rangle=\tilde G_{\mu\nu}$, since these quantities do not depend on $y$, and, for quantities which do not depend on the extra-coordinate (including constants) the Gaussian average coincides with the given quantity: $\left\langle P({\bf x})\right\rangle=P({\bf x})$.

\section{Gravitating Matter}\label{matter}

One may wonder under which conditions the solution (\ref{warp-f}), (\ref{5d-sol}) is still valid and, at the same time, 4D gravitational equations other than those for vacuum are obtained for the background metric (\ref{metric'}). In order to seek for an answer one is to add a matter piece $\propto\int d^5x\sqrt{|g_5|}\,{\cal L}_m(\chi,\nabla\chi,g_{AB})$ to the action (\ref{action}), where ${\cal L}_m$ is the Lagrangian density of the matter degrees of freedom $\chi$. The microscopic field equations obtained are

\be G_{AB}=\kappa_5^2 \left[T^{(m)}_{AB}+T^{(\vphi)}_{AB}\right],\;\Box_5\vphi=dV/d\vphi,\label{m-efe+kge}\ee where $\sqrt{|g_5|}\,T^{(m)}_{AB}=-2\,\delta\left(\sqrt{|g_5|}{\cal L}_m\right)/\delta g^{AB}$, is the matter stress-energy and $T^{(\vphi)}_{AB}$ is the one for the scalar field (\ref{phi-stress-energy-t}). In terms of the metric (\ref{metric'}) equations (\ref{m-efe+kge}) are transformed into the following system of equations:

\bea &&\frac{\kappa_5^2\,T^{(m)}_{\mu\nu}-\tilde G_{\mu\nu}}{a^2}=\left[3H'+6H^2+\frac{\kappa_5^2}{2}(\vphi'^2+2V)\right]\tilde g_{\mu\nu},\nonumber\\
&&\frac{\tilde R}{2a^2}+\kappa_5^2\,T^{(m)}_{55}=6H^2-\frac{\kappa_5^2}{2}\left(\vphi'^2-2V\right),\nonumber\\
&&T^{(m)}_{\mu 5}+\vphi_{,\mu}\vphi'=0,\;\vphi''+4H\vphi'=dV/d\vphi.\label{explicit-efe}\eea 

As seen, if one requires that $\vphi=\vphi(y)$ $\Rightarrow\vphi_{,\mu}=0$, then it has to be assumed that the 5D matter stress-energy tensor does not have mixed components: $T^{(m)}_{\mu 5}=0$. Besides, since we have assumed (\ref{warp-f},\ref{5d-sol}) is valid, then from (\ref{explicit-efe}) it follows that: $\tilde G_{\mu\nu}=\kappa_5^2\,T^{(m)}_{\mu\nu}$, $\kappa_5^2\,T^{(m)}_{55}=-\tilde R/2a^2$. These equations lead to the following conditions on the 5D matter's stress-energy tensor: 

\be T^{(m)}_{\mu 5}=0,\;T^{(m)}_{55}=g^{\mu\nu}T^{(m)}_{\mu\nu}/2.\label{conditions}\ee If these conditions are met, then standard 4D microscopic EFE-s: $\tilde G_{\mu\nu}=\kappa_5^2\,T^{(m)}_{\mu\nu}({\bf x},y)$, are recovered from Eq.(\ref{m-efe+kge}). These are to be regarded as a set of one-parametric partial differential equations. The continuous parameter $y$ is what makes the microscopic $T^{(m)}_{\mu\nu}$-s to behave as random quantities.

It is difficult to find a 5D stress-energy tensor derived from a Lagrangian which meets the conditions Eq.(\ref{conditions}) at once, notwithstanding, here we assume it is possible in principle and defer this subject for future work. A trivial example of a stress-energy tensor which meets (\ref{conditions}) is supplied by 4D matter in the form of radiation: $g^{\mu\nu}T^{(rad)}_{\mu\nu}=0$ $\Rightarrow$ $T^{(rad)}_{55}=0$. In this case we can write $T^{(rad)}_{\mu\nu}=\rho_{rad}\left(4u_\mu u_\nu+g_{\mu\nu}\right)/3$, where $\rho_{rad}=\rho_{rad}({\bf x})$ is the energy density of radiation measured by a 4D observer. Also, since $u_\mu=a\,\tilde u_\mu$, $u^\mu=\tilde u^\mu/a$, $g_{\mu\nu}=a^2\tilde g_{\mu\nu}$, then $T^{(rad)}_{\mu\nu}=a^2\tilde T^{(rad)}_{\mu\nu}=a^2\rho_{rad}\left(4\tilde u_\mu \tilde u_\nu+\tilde g_{\mu\nu}\right)/3$, where $\tilde T^{(rad)}_{\mu\nu}$ is the macroscopic stress-energy tensor of the radiation. The macroscopic Einstein's equations obtained after averaging the microscopic ones are $$\tilde G_{\mu\nu}=\kappa_5^2\left\langle T^{(rad)}_{\mu\nu}\right\rangle=8\pi G_N\,\tilde T^{(rad)}_{\mu\nu},\;8\pi G_N=\sqrt{2/3}\kappa_5^2.$$ Since $M^2_{\rm Pl}=\sqrt{3/2}/\kappa_5^2$, then, the mass hierarchy can not be addressed if minimal coupling of matter to gravity is assumed.

\begin{figure}[t!]
\includegraphics[width=3.5cm,height=3.5cm]{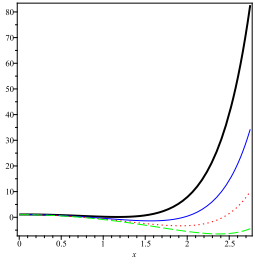}
\caption{The chameleon potential $V_{eff}$ (\ref{eff-cham-pot}) is shown for the following arbitrarily chosen values of the parameters: $V_0=1$, $\mu=1$, $\beta=0.5$, and $\tau=0.1$ - thick solid curve, $\tau=5\times 10^{-2}$ - thin solid line, $\tau=2.5\times 10^{-2}$ - dotted line, $\tau=10^{-2}$ - dashed curve. In general $\tau=\tau({\bf x})$ is a function of the spacetime point so that the 4D dynamics affects the position (and existence) of the minimum.}
\label{fig2}
\end{figure}

\subsection{Chameleon Coupling of Matter}

In this subsection, in order to be able to address the mass hierarchy issue within our approach, we shall allow for a non-minimal (chameleon) coupling of matter to the scalar field where, in harmony with effective string theory \cite{wands}, the strength of the coupling $\beta$ should be of order unity. In this case it is fundamental that the scalar field's potential $V(\vphi)$ be stable, since otherwise, there will be a serious problem with stabilization of the $y$-position of the Gaussian distribution function for the matter. 

Here we assume $\vphi\equiv\vphi({\bf x},y)=\vphi(y)+\phi({\bf x})$. Hence, since $G_{5\mu}=0$, $e^{5\kappa_5\beta(\vphi+\phi)}T^{(m)}_{5\mu}=-\vphi'\phi_{,\mu}\neq 0$. We shall consider a matter field $\chi$ which couples to the conformal metric in the way a chameleon does \cite{chameleon}: $\hat g_{AB}=e^{2\kappa_5\beta\vphi}\,g_{AB}$. Hence, the chameleonic matter piece action is given by $S_5^{(m)}=\int d^5x\sqrt{|\hat g_5|}\,{\cal L}_m(\chi,\nabla\chi,\hat g_{AB})$. The microscopic field equations (\ref{m-efe+kge}) are now replaced by: $G_{AB}=\kappa_5^2 \left[e^{3\kappa_5\beta\vphi}\,\hat T^{(m)}_{AB}+T^{(\vphi)}_{AB}\right]$, and the KG equation $\Box_5\vphi=dV/d\vphi-\kappa_5\beta\,e^{5\kappa_5\beta\vphi}\hat T^{(m)}$, where, $$\hat T^{(m)}_{AB}=-\frac{2}{\sqrt{|\hat g_5|}}\frac{\delta\left(\sqrt{|\hat g_5|}{\cal L}_m\right)}{\delta\hat g^{AB}}=e^{2\kappa_5\beta\vphi}\,T^{(m)}_{AB},$$ is the chameleonic matter stress-energy tensor with trace $\hat T_5^{(m)}=\hat g^{MN}\hat T^{(m)}_{MN}=T_5^{(m)}$, and we have taken into account the relationship $\delta g^{NM}=e^{2\kappa_5\beta\vphi}\delta_A^N\delta_B^M\delta\hat g^{AB}$. In terms of the stress-energy tensor $T^{(m)}_{AB}$ the resulting microscopic EFE and the KGE are:

\bea &&G_{AB}=\kappa_5^2 \left[e^{5\kappa_5\beta\vphi}\,T^{(m)}_{AB}+T^{(\vphi)}_{AB}\right],\nonumber\\
&&\Box_5\vphi=dV/d\vphi-\kappa_5\beta\,e^{5\kappa_5\beta\vphi}\,T_5^{(m)},\label{chameleon-efe}\eea respectively, or, if substitute the metric (\ref{metric'}) into (\ref{chameleon-efe})

\bea &&\kappa_5^2\,e^{5\kappa_5\beta(\vphi+\phi)} T^{(m)}_{\mu\nu}+\kappa_5^2\tilde T^{(\phi)}_{\mu\nu}-\tilde G_{\mu\nu}=\nonumber\\
&&\;\;\;\;\;\;\;\;\;\;\;\;\;\;\;g_{\mu\nu}\left[3H'+6H^2+\frac{\kappa_5^2}{2}\left(\vphi'^2+2V(\vphi)\right)\right],\nonumber\\
&&\frac{\tilde R-\kappa_5^2(\tilde\nabla\phi)^2}{2a^2}+\kappa_5^2\,e^{5\kappa_5\beta(\vphi+\phi)} T^{(m)}_{55}=\nonumber\\
&&\;\;\;\;\;\;\;\;\;\;\;\;\;\;\;\;\;\;\;\;\;\;\;\;\;\;\;\;\;\;\;\;\;\;6H^2-\frac{\kappa_5^2}{2}\left(\vphi'^2-2V(\vphi)\right),\nonumber\\
&&\frac{\tilde\Box\phi}{a^2}+\kappa_5\beta\,e^{5\kappa_5\beta(\vphi+\phi)} T_5^{(m)}=\frac{dV}{d\vphi}-\vphi''+4H\vphi',\label{cham-feqs}\eea where $\tilde T^{(\phi)}_{\mu\nu}=\phi_{,\mu}\phi_{,\nu}-\tilde g_{\mu\nu}(\tilde\nabla\phi)^2/2$ is the stress-energy tensor for the (massless) scalar field's component $\phi=\phi({\bf x})$. Besides, the consistency condition $T^{(m)}_{55}=T^{(m)}_4/2$ ($T^{(m)}_4\equiv g^{\mu\nu}T^{(m)}_{\mu\nu}$) leads to $T^{(m)}_5=3T^{(m)}_4/2$. Here we do not care about a concrete realization of a matter stress-energy tensor which fulfills the above conditions and just assume it can be found. The trivial case which meets these conditions is the radiation matter field.

Under the ansatz (\ref{warp-f}) the following 5D solution is found:

\bea &&\vphi^\pm(y)=\pm\sqrt{3\mu^2/\kappa_5^2}\;y+\vphi_0,\nonumber\\
&&V(\vphi)=V_0-2\mu^2(\vphi-\vphi_0)^2,\label{chameleon-sol}\eea where $V_0=3\mu^2/2\kappa_5^2$, and $\vphi_0$ is a non-vanishing integration constant since, under the chameleon coupling, the mirror symmetry is not a symmetry of the field equations any more. The above 5D solution is consistent with the 4D microscopic EFE-s:

\bea &&\tilde G_{\mu\nu}=\kappa_5^2\,e^{5\kappa_5\beta(\vphi+\phi)} T^{(m)}_{\mu\nu}+\kappa_5^2\tilde T^{(\phi)}_{\mu\nu},\nonumber\\
&&\tilde\Box\phi=-3\kappa_5\beta\,e^{5\kappa_5\beta(\vphi+\phi)}\,a^2\,T^{(m)}_4/2.\label{4d-chameleon-efe}\eea 

Even if the potential $V$ in (\ref{chameleon-sol}) is unstable, what really matters to the dynamics of $\vphi({\bf x},y)$ is not $V$ itself, but the chameleon potential \cite{chameleon} 

\be V_{eff}=V_0-2\mu^2(\vphi-\vphi_0)^2+\tau\,e^{5\kappa_5\beta\vphi},\label{eff-cham-pot}\ee where $\vphi=\vphi(y)$ is given by (\ref{chameleon-sol}), and we have defined $\tau=\tau({\bf x})=-3\,e^{5\kappa_5\beta\phi}T^{(m)}_4/10$. If $\tau\geq 0$ ($T^{(m)}_4<0$) this potential is an extremum at $\vphi_*$ which solves the non-algebraic equation: $5\kappa_5\beta\tau\exp{(5\kappa_5\beta\vphi_*)}=4\mu^2(\vphi_*-\vphi_0)$. If $\vphi_*>\vphi_0+1/5\kappa_5\beta$ the extremum is a minimum at $\vphi_*$. Notice that, if assume $\beta$ is a non-negative constant, then only for the positive branch of the solution (\ref{chameleon-sol}) the minimum of $V$ can be found. In what follows we shall consider only the positive branch of (\ref{chameleon-sol}). The value $\vphi_*$ picks a position in the extra-space $y_*=\sqrt{\kappa_5^2/3\mu^2}\,(\vphi_*-\vphi_0)$, which meaning will be clear below. In the figure \ref{fig2} the form of the potential (\ref{eff-cham-pot}) is shown for arbitrarily chosen values of the parameters $V_0$, $\mu$, and $\beta$. Different constant values have been also assigned to the function $\tau$. However, since this is a function of the spacetime point, the 4D dynamics affects the existence and position of the minimum of the potential and, hence, the value $y_*$.

\begin{figure}[t!]
\includegraphics[width=4cm,height=3.5cm]{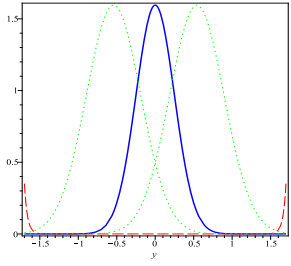}
\caption{The 4D graviton's probability density $\Psi^2_0(y)$ (solid line), the probability density of the massive states $\Psi^2_m(y)$ (dashed line), and the shifted probability density distribution $\Psi^2_{\rm shift}(y)$ (doted line), are shown. The following values of the parameters have been chosen: $(\Delta,m,\beta)=(0.5,0.01,0.5)$.}
\label{fig3}
\end{figure}

\subsection{Mass Hierarchy}

The chameleon coupling shifts the maximum of the probability density distribution for the matter degrees of freedom with respect to the gravitational ones. To see this let us to average the first term in the RHS of Einstein's equations in (\ref{4d-chameleon-efe}), 

\bea &&\kappa_5^2\left\langle e^{5\kappa_5\beta(\phi+\vphi)}\,T^{(m)}_{\mu\nu}\right\rangle=\nonumber\\
&&\;\;\;\;\;\;\;\kappa_5^2\,e^{5\beta\kappa_5\left(\phi+\vphi_0+\frac{15\beta}{8\kappa_5}\right)}\int_{-\infty}^{\infty} T^{(m)}_{\mu\nu}\,\Psi^2_{\rm shift}\,dy,\label{shift-ave}\eea where $\Psi^2_{\rm shift}(y)=(\sqrt{2/\pi}/\Delta)\,e^{-2(y-y_0)^2/\Delta^2}=\Psi^2_0(y-y_0)$, is the 'shifted' probability distribution function with $y_0=5\sqrt{3}\beta\Delta/4$. As it has been clearly illustrated in Fig.\ref{fig3}, it is apparent that, while the probability density function for the gravitational degrees of freedom $\Psi^2_0(y)$ in Eq.(\ref{gauss-d}) is peaked at the origin $y=0$, the one for the matter degrees of freedom $\Psi^2_{\rm shift}$ is peaked at $y=y_0$, i. e., it is shifted in the extra-direction.

Let us to do some juggling with the free parameters. First we write the integration constant $\vphi_0$ in terms of the values at the minimum of the effective chameleon potential $\vphi_0=\vphi_*-\sqrt 3\,\mu\,y_*/\kappa_5$ and substitute in Eq.(\ref{shift-ave}). Then we identify the position $y_0$ of the center of the shifted distribution function ($\Psi^2_{\rm shift}$) with the position $y_*$ picked up by the minimum of $V_{eff}$: $y_0=y_*$. This choice forces the relationship $\beta=(4/5\sqrt 3)\,y_*/\Delta$. Since the integration constant $\vphi_0$ is arbitrary, we can make yet another assumption on the constants: $\vphi_*=\sqrt 3\,\mu\,y_*/2\kappa_5$ $\Leftrightarrow\;\vphi_0=-\vphi_*$, which greatly simplifies the resulting expressions. After the above choice of the free parameters Eq.(\ref{shift-ave}) can be written in the following way: $$\kappa_5^2\left\langle e^{5\kappa_5\beta(\phi+\vphi)}T^{(m)}_{\mu\nu}\right\rangle=\kappa_5^2\,e^{5\kappa_5\beta\phi}\left\langle T^{(m)}_{\mu\nu}\right\rangle_*,$$ where ''$\left\langle \right\rangle_*$'' means averaging over the shifted Gaussian distribution $\Psi^2_*(y)=\Psi^2_0(y-y_*)$. The 4D macroscopic EFE-s -- the ones which dictate the gravitational laws a 4D observer like us measures -- can be obtained through Gaussian averaging of the EFE-s (\ref{4d-chameleon-efe}). We obtain: $$\tilde G_{\mu\nu}=\kappa_5^2\,e^{5\kappa_5\beta\phi}\left\langle T^{(m)}_{\mu\nu}\right\rangle_*+\kappa_5^2\,\tilde T^{(\phi)}_{\mu\nu}.$$ 

Let us, for definiteness, to assume $T^{(m)}_{\mu\nu}$ in the above equation can be written as $T^{(m)}_{\mu\nu}({\bf x},y)=a^2(y)\,\tilde T^{(m)}_{\mu\nu}({\bf x})$ $\Rightarrow\;T^{(m)}_4=\tilde T^{(m)}_4$ (see the former section). Hence $$\left\langle T^{(m)}_{\mu\nu}\right\rangle_*=\sqrt{2/3}\;e^{-\frac{2y^2_*}{3\Delta^2}}\,\tilde T^{(m)}_{\mu\nu}.$$ Accordingly, the averaged EFE-s $$\tilde G_{\mu\nu}=8\pi G_N(e^{5\kappa_5\beta\phi}\,\tilde T^{(m)}_{\mu\nu}+\tilde T^{(\phi)}_{\mu\nu}),$$ can be obtained if the following relationship between ''Newton's gravitational constant'' and the 5D gravitational coupling takes place:  $$8\pi G_N=\sqrt{2/3}\,e^{-\frac{25}{8}\beta^2}\kappa_5^2\;\Rightarrow\;M_{\rm Pl}^2=\sqrt\frac{3}{2}\,e^{\frac{25}{8}\beta^2}\frac{1}{\kappa_5^2}.$$

If assume that the fundamental 5D scale $1/\kappa_5\sim\text{TeV}$, and take into account that the effective Planck scale $M_{\rm Pl}\cong 10^{19}\text{GeV}$, then, from the above equation one finds $$\beta^2=\frac{16}{75}\left(\frac{y_*}{\Delta}\right)^2\simeq\frac{256}{25}\ln(10)\;\Rightarrow\;\beta\simeq 4.88\,.$$ Hence, we do not need a large value of the constant parameter $\beta$ to explain the large hierarchy between the mass scales. Recall that, in harmony with string theory, $\beta\sim 1$ \cite{wands,chameleon}.

\section{Discussion and Conclusion}

In the present paper we have addressed the issue of how to compute in an unambiguous way effective four-dimensional quantities within thick brane contexts \cite{mounaix,quiros-matos}. Our proposal rests upon the following assumptions: i) the background geometry is a warped one with a Gaussian warp factor (\ref{warp-f}), and ii) the extra-coordinate $y$ is regarded as a continuous randomly distributed parameter. The resulting picture supports the following probabilistic interpretation. Fields which depend on the extra-coordinate are also randomly distributed (we call them as microscopic fields), their probability distribution density is a Gaussian with variance $\sigma=\Delta/2$, where $\Delta$ is the width of the brane. The corresponding 4D measurable quantities coincide with their Gaussian averages.

As in RS braneworlds \cite{rs,rs2} the warped background geometry warrants that a bound 4D graviton exists and is capable of reproducing standard 4D laws of gravity in the brane. This time, however, due to the well-shaped form of the analog quantum mechanics potential $U_{QM}$ (see Fig.\ref{fig1}), even if the excited KK states form a continuum, there exists an effective mass gap between the excited KK modes and the ground massless state. This is explained by the fact that the probability density of the KK states is a vanishing minimum at the brane location and, as we depart from the brane, it grows slower for the lighter KK modes. This means that the heavier KK excitations are closer to the brane than the lighter ones and, hence, can interact with the graviton with a greater probability. However, low energy states living in the thick brane need of an energy of the order of the mass of the KK modes to excite them. The interplay between probability of interaction and mass of the KK states is what generates the effective mass gap. 

Another ingredient of our approach is related with the way measurable 4D quantities are obtained. Here the effective picture is generated by averaging the microscopic equations and quantities over the Gaussian distribution for the random parameter $y$ to get the corresponding macroscopic equations and quantities, i. e., the ones a 4D observer living in the thick brane is capable of measuring. We want to underline that the RS solution \cite{rs2} can not be recovered from our model under any circumstances, i. e., the latter does not represent a thick brane generalization of the RS model as it is customary \cite{csaba,arias,polacos}. It is just a different phenomenological scenario. The fact that the solution (\ref{warp-f},\ref{5d-sol}) supports a smooth thick brane configuration is a direct consequence of the (linear) gravitational content of our model and of our choice of the way observable quantities are defined.

In order to be able to explain the large hierarchy between the TeV and Plack scales in our approach one needs to assume a non-minimal (chameleonic) interaction between the matter degrees of freedom living in the brane and the 5D scalar field. An interesting picture arises. The resulting probability distribution function for matter, which is shaped by $\Psi^2_*$, is shifted in the extra-space with respect to the one for gravity, which is shaped by $\Psi^2_0$: $\Psi^2_*=\Psi^2_0(y-y_*)$, where the position $y_*$ of (the center of) the matter probability distribution is determined by the minimum of the effective potential for $\vphi({\bf x},y)=\vphi(y)+\phi({\bf x})$: $V_{eff}=V_0-2\mu^2(\vphi-\vphi_0)^2+\tau\,e^{5\kappa_5\beta\vphi}$, whenever the minimum exists. Besides, since $\tau=-3\,e^{5\kappa_5\beta\phi}T^{(m)}_4/10=\tau({\bf x})$, the 4D spacetime dynamics of matter affects the 5D dynamics through modifying the position $y_*$ of the matter probability distribution function. If assume, for instance, 4D matter in the form of a perfect fluid with energy density $\rho({\bf x})$ and pressure $p=(\gamma-1)\rho$ ($\gamma$ is the barotropic parameter), then $\tau({\bf x})\propto(4-3\gamma)\,e^{5\kappa_5\beta\phi({\bf x})}\,\rho({\bf x})$. For larger values of $\rho$ the position $y_*$ of the matter probability distribution function $\Psi^2_*$ -- the one which shapes the thick brane where the matter is trapped -- is closer to $y=0$ where the graviton's probability distribution function is a maximum, while for smaller $\rho$-s it is farther away (see Fig.\ref{fig2}). For a relativistic fluid with $\gamma=2$ (stiff-matter), since $\tau<0$, the minimum does not exist and the position of the matter probability distribution is not stabilized at all. Within a cosmological setting this would mean that, close to the big bang, when relativistic matter is expected to dominate the cosmic dynamics, the position in the extra-space of the thick brane where matter lives (the one shaped by $\Psi^2_*$) is not stabilized. As long as non-relativistic matter starts dominating, a minimum of $V_{eff}$ is found and the matter brane's position is stabilized at some $y_*$. As the expansion further proceeds and the matter energy density dilutes, the position $y_*$ of the matter brane gets farther and farther away from the brane at $y=0$ where the massless graviton lives, which is shaped by $\Psi^2_0$. This entails that the strength of gravitational interactions felt by matter confined to the thick brane at $y_*$ weakens with the cosmic expansion. This happens after the radiation domination epoch ($\tau=0$) where the scalar field's potential drives unstable 5D dynamics. Before that stage, as said, perhaps relativistic matter dominated so that stabilization of the matter's thick brane position was not achieved neither. A more careful study, including detailed investigation of the 4D cosmic dynamics within the present approach, will lead to new interesting features not explored in the present paper.  

The authors acknowledge useful comments by A Ahmed. I Q thanks the mathematics department at CUCEI, Guadalajara University, and SNI of Mexico for support. The work of T M was partially supported by CONACyT Mexico grant 49865-E.

\end{document}